\documentclass[11pt]{article}
\usepackage[left=1in,top=1in,right=1in,bottom=1in,head=0in,nofoot]{geometry}

\usepackage{setspace,url,bm,amsmath} 
\usepackage{scalerel}

\usepackage{xcolor}
\usepackage{multirow}
\usepackage{arydshln}
\usepackage{mathtools}
\usepackage{subcaption}
\usepackage{titlesec} 
\titlelabel{\thetitle.\quad} 
\titleformat*{\section}{\bf\Large\center}

\usepackage{authblk}
\usepackage{float}
\usepackage{graphicx} 
\usepackage{bbm}
\usepackage{latexsym}
\usepackage{caption}
\usepackage[margin=20pt]{subcaption}
\usepackage{enumerate}
\graphicspath{ {./Graphs/} }

\usepackage{tcolorbox}
\usepackage{pdflscape} 

\usepackage{amsthm}
\usepackage{amssymb}
\usepackage{amsmath}
\usepackage{color}

\usepackage{comment}
\theoremstyle{definition}
\newtheorem{assumption}{Assumption}
\newtheorem*{theorem*}{Theorem}
\newtheorem{theorem}{Theorem}
\newtheorem*{rmk*}{remark}
\newtheorem{proposition}{Proposition}
\newtheorem{lemma}{Lemma}

\newtheorem{remark}{Remark}
\newtheorem{corollary}{Corollary}
\newtheorem*{corollary*}{Corollary}

\RequirePackage[normalem]{ulem}

\usepackage{natbib} 
\bibpunct{(}{)}{;}{a}{}{,} 

\usepackage{etoolbox} 
\apptocmd{\sloppy}{\hbadness 10000\relax}{}{} 

\usepackage{multibib}
\newcites{sec}{References}

\usepackage{color}
\usepackage{listings}
\usepackage{hyperref}
\usepackage{ulem} 

\usepackage{enumitem}
\usepackage{booktabs}
\usepackage{lscape}
\usepackage{pgfplots}
\usetikzlibrary{intersections}
\usepgfplotslibrary{fillbetween}

\def \E {\mathbb{E}}

\newcommand{\indep}{\perp \!\!\! \perp}

\usepackage{pgf,tikz}
\usepackage{tikz}
\usetikzlibrary{shapes,arrows,positioning,calc}

\usepackage{wasysym}

\allowdisplaybreaks

\usepackage{xr}
\makeatletter
\newcommand*{\addFileDependency}[1]{
  \typeout{(#1)}
  \@addtofilelist{#1}
  \IfFileExists{#1}{}{\typeout{No file #1.}}
}
\makeatother

\newcommand*{\myexternaldocument}[1]{%
    \externaldocument{#1}%
    \addFileDependency{#1.tex}%
    \addFileDependency{#1.aux}%
} 
\myexternaldocument{suppl_v2}

\begin{document}

\singlespacing

\title{\bf Partial Identification under Imperfect Compliance via Cross-World Correlation Restriction} 

\author[1]{Zhaoqing Tian}
\author[1,2,3]{Lin Liu}
\author[4]{Yue Liu\footnote{Corresponding authors: liuyue\_stats@ruc.edu.cn and pengwu@btbu.edu.cn.}}
\author[5]{Peng Wu$^*$}

\affil[1]{\small School of Mathematical Sciences, Shanghai Jiao Tong University, Shanghai, China} 
\affil[2]{\small Institute of Natural Sciences, MOE-LSC, SJTU-Yale Joint Center for Biostatistics and Data Science, Shanghai Jiao Tong University, Shanghai, China} 
\affil[3]{\small Research Center, Pu'er People's Hospital, Yunnan, China} 
\affil[4]{\small Center for Applied Statistics and School of Statistics, Renmin University of China, Beijing,  China} 
\affil[5]{\small School of Mathematics and Statistics, Beijing Technology and Business University, Beijing, China}  

\date{}

\maketitle

\begin{abstract} 
This article develops a unified partial identification approach for the average treatment effect (ATE) and the local average treatment effect (LATE) in randomized experiments with imperfect compliance, incorporating the cross-world correlation between potential treatment statuses. Even under random assignment and  exclusion restriction, the correlation is unidentified. However, given the two identified marginals, it uniquely determines the relative proportions of the four principal strata, with monotonicity corresponding to a special case. 
We derive explicit sharp bounds for both the ATE and LATE at a fixed correlation and extend the results to prespecified correlation ranges, providing a sensitivity analysis that incorporates substantive information about compliance behavior. For the ATE, the lower bound decreases and the upper bound increases with the correlation. Although imposing monotonicity does not tighten the conventional Balke--Pearl bounds, we show that restricting the correlation can yield 
tighter bounds, even when defiers remain possible. For the LATE, the sharp bounds may not vary monotonically with the correlation, and we establish necessary and sufficient conditions for point identification of the LATE at any feasible correlation value, with the conventional Wald ratio identification result under monotonicity as a special case. In addition, we derive a breakdown point for the correlation that determines when the Wald ratio falls outside the sharp identified set for the LATE. Applications to four empirical studies illustrate how cross-world correlation restrictions tighten the sharp bounds for the ATE and LATE and assess the breakdown of the Wald ratio.  
\end{abstract}


\newpage
 
\onehalfspacing

\section{Introduction} \label{intro}

Instrumental variable (IV) methods provide a powerful framework for causal inference when actual treatment differs from treatment assignment (instrument), as commonly occurs in randomized experiments and encouragement designs~\citep[see e.g.,][]{Bloom1984,Robins01011994,hernan2006instruments,baiocchi2014instrumental,imbens2014instrumental,AngristHull2023}.  In such settings, the causal effect of the instrument may differ from that of the treatment actually received.
IV methods exploit instrument-induced variation in treatment receipt to identify or bound causal effects.
Point and partial identification typically rely on the random assignment and exclusion restriction, sometimes supplemented by assumptions such as monotonicity, which restrict how potential treatment receipt can vary across 
 instrument assignments.  

In this article, we develop a unified partial identification approach for the average treatment effect (ATE) and the local average treatment effect (LATE) in randomized experiments with imperfect compliance. Let $T(0)$ and $T(1)$ denote the potential treatment statuses under instrument assignments $Z=0$ and $Z=1$, respectively. Our approach uses their Pearson correlation, $\rho=\operatorname{Corr}(T(0),T(1))$, as the sensitivity parameter.  
We refer to \(\rho\) as the cross-world correlation because it captures the dependence between the potential treatment statuses under the two hypothetical worlds. This terminology is analogous to the cross-world assumption in \citet{Bodik-etal2025}. 
This correlation provides a parsimonious and interpretable way to incorporate information about latent compliance behavior. Specifically, the joint values of $(T(0),T(1))$ define four principal strata for binary treatment: always-takers, never-takers, compliers, and defiers. Given the marginal treatment probabilities, each feasible value of $\rho$ uniquely determines the proportions of these four strata. The conventional monotonicity assumption, $T(1)\geq T(0)$, corresponds to the upper endpoint of the feasible correlation range, at which the proportion of defiers is zero. 

Our article makes three main contributions. First, for binary outcomes, we derive explicit sharp bounds for the ATE under a fixed cross-world correlation and extend these results to prespecified correlation ranges. Previous studies~\citep{Balke01091997, BaiEtAl2026} have established that, under random assignment and the exclusion restriction, imposing monotonicity does not sharpen the Balke--Pearl bounds\footnote{Balke--Pearl bounds are sharp bounds for the ATE under random assignment and the exclusion restriction alone.} for the ATE. We extend this result by showing that restricting the cross-world correlation can nevertheless yield strictly tighter bounds, even when defiers remain possible. Moreover, we establish an explicit threshold such that the ATE bounds become strictly tighter whenever the upper endpoint of the prespecified correlation range is below this threshold. 
This result highlights a somewhat counterintuitive feature: although restricting the cross-world correlation allows defiers to remain possible and thus imposes a less restrictive condition on compliance behavior than monotonicity, it can nevertheless yield strictly tighter ATE bounds.

Second, we derive explicit sharp bounds for the LATE under fixed cross-world correlations and characterize the sharp identified set under prespecified correlation ranges.
Unlike the ATE bounds, which vary monotonically with $\rho$, the LATE bounds can exhibit nonmonotone behavior: the lower bound may first decrease and then increase, while the upper bound may first increase and then decrease. Consequently, the sharp identified sets at different values of $\rho$ need not be nested. 
Moreover, we establish necessary and sufficient conditions for point identification of the LATE at any feasible value of \(\rho\), with the conventional point identification of the LATE by the Wald ratio under monotonicity~\citep{ImbensAngrist1994,Angrist01061996,HullKolesarWalters2022} as a special case.
These results indicate the distinct implications of cross-world correlation restrictions for the partial identification of the ATE and LATE.

Third, we explore whether the Wald ratio
falls within the sharp identified set for the LATE under restrictions on $\rho$.  
We establish an explicit threshold such that the Wald ratio is excluded from the sharp identified set whenever the upper endpoint of the prespecified correlation range is below this threshold.
We further compare this threshold with the one for improving the ATE bounds, showing that the correlation restriction can first sharpen the ATE bounds while retaining the Wald ratio as a possible LATE value, and then exclude the Wald ratio as the restriction becomes more stringent.
Thus, our approach provides a unified approach to partial identification of the ATE and LATE and to assessing the breakdown of the Wald ratio under cross-world correlation restriction. 
Furthermore, we illustrate the proposed approach through four empirical applications to randomized studies with imperfect compliance.


Our work relates to three strands of the IV literature. The first concerns the partial identification of treatment effects under random assignment and the exclusion restriction. 
Following the early nonparametric bounds of \citet{Robins1989} and \citet{Manski1990}, \citet{Balke01091997} derived sharp bounds for the ATE in the binary IV model.
\citet{kitagawa2021identification} characterized the sharp identified set for the marginal potential outcome distributions and extended the Balke--Pearl ATE bounds from binary outcomes to more general bounded scalar outcomes.
\citet{RichardsonRobins2010,Richardson&Robins2014} and \citet{SwansonEtAl2018} reviewed the sharp bounds for ATE under different versions of the random assignment and exclusion restriction assumptions. 
More recently, \citet{LevisEtAl2025} incorporated  covariates to tighten ATE bounds and develop influence-function-based estimation and inference methods. 
\citet{kedagni2024robust} derived bounds on the LATE   
and used plug-in estimators and an intersection-bound inference procedure to construct confidence bands.  
The second strand imposes monotonicity and examines how this additional assumption sharpens the partial identification of treatment effects under random assignment and the exclusion restriction. 
Under these assumptions, \citet{ImbensAngrist1994} and \citet{Angrist01061996} showed that the LATE is point identified by the Wald ratio, and \citet{ImbensRubin1997Distributions} showed how to identify and estimate the marginal distributions of potential outcomes among compliers. For the ATE, prior work~\citep{Balke01091997,SwansonEtAl2018,BaiEtAl2026} has shown that imposing monotonicity does not tighten the Balke--Pearl bounds in the binary IV model. 


The third strand relaxes monotonicity to allow for defiers. One line of work achieves point identification under alternative assumptions. 
\citet{deChaisemartin2017} and \citet{small2017instrumental} showed that the Wald ratio can point identify several variants of ATEs under certain conditions even when defiers are present.
\citet{dahl2023never} showed that the ATE for compliers and defiers are separately point identified under a weaker form of monotonicity that allows for defiers. 
The other line of work examines how conclusions about treatment effects vary with restrictions on the defier share. 
\citet{SwansonHernan2018}, \citet{Huber2014}, and \citet{noack2021sensitivity} used the defier share as a sensitivity parameter to assess how varying this share affects conclusions about the sign and magnitude of the LATE.
For the ATE, related work studied how the lower and upper bounds change under different restrictions on the defier share~\citep{noack2021sensitivity,Yap2025}.
More recently, \citet{picchetti2026sensitivity} derived ATE and LATE bounds under joint relaxations of monotonicity and random assignment (or instrument independence) and developed estimation and inference methods. 
Our approach provides a unified framework for partial identification of the ATE and LATE, using the cross-world correlation \(\rho\) as the common sensitivity parameter. It encompasses established results, including the Balke--Pearl bounds for the ATE and identification of the LATE by the Wald ratio under monotonicity, while yielding three new insights: (i) restrictions on $\rho$ can strictly sharpen the Balke--Pearl bounds even when defiers remain possible; (ii) necessary and sufficient conditions characterize point identification of the LATE at a fixed $\rho$; and (iii) an explicit ordering of the thresholds for sharpening the ATE bounds and excluding the Wald ratio from the LATE identified set. These results reveal the distinct yet interconnected implications of cross-world correlation restrictions for the ATE and LATE.

The remainder of the article is organized as follows. Section~\ref{sec2} introduces the potential-outcomes notation, the identifying assumptions, and the conventional IV benchmarks. Section~\ref{sec3} develops the cross-world correlation framework and establishes the sharp bounds for the ATE and LATE, together with the corresponding identification and Wald compatibility results. 
Section~\ref{sec4}  presents four empirical applications. Section~\ref{sec5} concludes with a discussion. 
Technical proofs and additional analysis are provided in the Supplementary Material.

\section{Setup}  \label{sec2}
We consider a randomized experiments with imperfect compliance.  
Let \(Z\in\{0,1\}\) denote the treatment assignment (instrument), and let \(T\in\{0,1\}\) denote the treatment actually received, where 1 and 0 represent treatment and control, respectively. Let \(Y\in\{0,1\}\) denote the binary outcome of interest. 
We use the potential outcomes framework to define causal effects. Let \(T(z)=T(Z=z)\) for \(z=0,1\) denote the treatment that would be received under assignment \(Z=z\), and let \(Y(t,z)=Y(T=t,Z=z)\) for \(t,z=0,1\) denote the potential outcome under treatment status \(T=t\) and assignment \(Z=z\). 
Under the stable unit treatment value assumption, which rules out multiple versions of treatment and interference among individuals, the observed treatment and outcome satisfy \(T=T(Z)\) and \(Y=Y(T,Z)\), respectively. That is, the observed treatment and outcome equal their corresponding potential values under the realized assignment and treatment. 

Individuals can be classified into four principal strata according to their potential treatment statuses under the two instrument assignment conditions: always-takers \((T(0)=1,T(1)=1)\), never-takers \((T(0)=0,T(1)=0)\), compliers \((T(0)=0,T(1)=1)\), and defiers \((T(0)=1,T(1)=0)\). These strata represent distinct patterns of latent compliance behavior. 
Throughout, we maintain the standard  assumptions of exclusion restriction and random assignment. 

\begin{assumption}[Exclusion Restriction] \label{assump1} $Y(t, 0) = Y(t, 1)$ for $t = 0, 1$.    
 \end{assumption}
 
 Assumption \ref{assump1} states that, for a given treatment status $t$, changing the instrument assignment from $Z=0$ to $Z=1$ does not affect the potential outcome. It rules out a direct effect of the instrument on the outcome and requires the instrument to affect the outcome only through the treatment received. Accordingly, we write $Y(t,z)=Y(t)$.  

\begin{assumption}[Random Assignment] \label{assump2}
$Z \indep \{T(0), T(1), Y(0), Y(1)\}$, and $ 0 < \Pr(Z=1) < 1$. 
\end{assumption}	

Assumption \ref{assump2} rules out unmeasured confounding between \(Z\) and \((T,Y)\), while allowing for unmeasured confounding between \(T\) and \(Y\). 
The condition \(0<\Pr(Z=1)<1\) ensures that both assignment levels occur with positive probability. This assumption is justified when the assignment is determined by an external randomization mechanism.  

Under noncompliance, suppose that 
$\Pr(T(0)=0,T(1)=1)>0.$  
Two commonly studied causal estimands are the average treatment effect (ATE) and the local average treatment effect (LATE), defined as 
\begin{align*}
\tau &= \E[Y(1)-Y(0)],\\ 
\tau_c &= \E[Y(1)-Y(0)\mid T(0)=0,T(1)=1].
\end{align*}
The LATE represents the average treatment effect among compliers, whose treatment receipt is induced by the instrument.  

We introduce the following notation, which will be used throughout the paper. Let
$$
p_{yt\mid z} 
\equiv \Pr(Y=y,T=t\mid Z=z), 
\qquad y,t,z\in\{0,1\}.
$$
$$
\lambda_z=\Pr(T=1\mid Z=z),
\qquad
\mu_z=\Pr(Y=1\mid Z=z),
$$
and let $\Delta_T=\lambda_1-\lambda_0$,  $\Delta_Y=\mu_1-\mu_0,$ and  
    \[   \gamma = \sqrt{ \lambda_0 (1 - \lambda_0) \lambda_1 (1- \lambda_1) }. \] 
We assume that \(Z\) is associated with \(T\), as required by the IV relevance condition, so that \(\Delta_T\neq0\). Without loss of generality, we take \(\Delta_T>0\) throughout, since interchanging the labels \(Z=0\) and \(Z=1\) reverses the sign of \(\Delta_T\) when it is negative. This restriction also facilitates comparison with the results under the additional monotonicity assumption \(T(1)\geq T(0)\), which requires \(\Delta_T\geq 0\) and is incompatible with the observed data when \(\Delta_T<0\). 

The following are standard identification results.
The ATE is not point identified under Assumptions~\ref{assump1}--\ref{assump2}.   \citet{Balke01091997} derived sharp bounds for the ATE using linear programming and showed that these bounds admit closed-form expressions; see Supplementary Material S1 for a brief overview.  
In addition, under Assumptions~\ref{assump1}--\ref{assump2} and the monotonicity assumption, the LATE is point identified by the Wald ratio :
$w = \Delta_Y/\Delta_T$.   
\section{Partial Identification under Cross-World Correlation Restrictions} \label{sec3}
In this section, we develop sharp partial identification bounds for both the ATE and LATE under Assumptions~\ref{assump1}--\ref{assump2} and restrictions on the cross-world correlation $\rho=\operatorname{Corr}(T(0),T(1))$. We first examine how $\rho$ determines the proportions of the four principal strata and its relationship to the monotonicity assumption. We then derive sharp bounds for the ATE and LATE under fixed values and prespecified ranges of $\rho$. Finally, we establish threshold conditions that characterize when restrictions on $\rho$ strictly tighten the ATE bounds and when they exclude the Wald ratio from the sharp identified set for the LATE.  

\subsection{Cross-World Correlation and Principal Strata}
\label{sec3-1}

We use the cross-world correlation $\rho$ as a sensitivity parameter to characterize the dependence between the potential treatment statuses. This parameter has several useful features. Given the marginal treatment probabilities, it uniquely determines the proportions of the four principal strata, with the conventional monotonicity assumption corresponding to the upper endpoint of its feasible range. Moreover, restrictions on $\rho$ can incorporate substantive knowledge about latent compliance behavior without requiring monotonicity.
We discuss these properties in detail below. 

\emph{First}, the cross-world correlation  $\rho$ is directly related to the proportions of the four principal strata under Assumption \ref{assump1}. Let 
	\[       \pi_{ij} =  \Pr( T(0) = i,  T(1) = j ), \quad \text{for } i, j \in \{0, 1\}.       \]
These proportions are not identified under Assumption~\ref{assump1} alone. However, for a given feasible value of $\rho$\footnote{We restrict $\rho$ to the feasible range implied by Assumption~\ref{assump1}; see equation~\eqref{eq2}.}, they are identifiable.  Specifically, writing $\pi_{ij}(\rho)$ to emphasize their dependence on $\rho$, we have 
\begin{equation}  \label{eq:stratum-proportions-rho}  
\begin{aligned}
  \pi_n(\rho)=\pi_{00}(\rho)
  &=(1-\lambda_0)(1-\lambda_1)+\rho \gamma,\\
  \pi_c(\rho)=\pi_{01}(\rho)
  &=(1-\lambda_0)\lambda_1-\rho  \gamma,\\
  \pi_d(\rho)=\pi_{10}(\rho)
  &=\lambda_0(1-\lambda_1)-\rho \gamma,\\
  \pi_a(\rho)=\pi_{11}(\rho)
  &=\lambda_0\lambda_1+\rho \gamma, 
\end{aligned}
\end{equation}
where $\pi_n(\rho)$, $\pi_c(\rho)$, $\pi_d(\rho)$, and $\pi_a(\rho)$ denote the proportions of never-takers, compliers, defiers, and always-takers, respectively. 
Thus, specifying $\rho$ resolves the uncertainty about the distribution of the four principal strata left by Assumption~\ref{assump1}. 

\emph{Second}, the cross-world correlation also provides a direct characterization of the conventional monotonicity assumption in the binary IV models. For completeness and to facilitate comparison with existing results, we state this assumption explicitly. 
\begin{assumption}[Monotonicity]\label{assump3}
    \(T(1)\geq T(0)\) almost surely.
\end{assumption}  
As noted by \citet{wu2026quantifying},  for binary treatment,  
 the feasible range of $\rho$ need not coincide with $[-1,1]$. Under Assumption \ref{assump1} and $\Delta_T > 0$, the Fr\'{e}chet--Hoeffding lower and upper bounds for $\rho$ are given as:  
    \begin{equation} \label{eq2} 
    L_\rho ={} - \frac{ \min\big\{ (1-\lambda_0)(1-\lambda_1), \lambda_0 \lambda_1\big\} }{\gamma},  \quad 
    U_\rho ={}  \frac{ \lambda_0 (1-\lambda_1) }{\gamma}.
    \end{equation}  
    Both endpoints are attainable and correspond to boundary configurations of the principal strata.
 Specifically, $\rho=L_\rho$ if and only if either $\pi_n=0$ or $\pi_a=0$, whereas $\rho=U_\rho$ if and only if $\pi_d=0$. The latter condition rules out defiers and is therefore equivalent to monotonicity. This relationship is formalized in the following proposition.

\begin{proposition} \label{prop1}
Under Assumption~\ref{assump1} and \(\Delta_T>0\), Assumption~\ref{assump3} holds if and only if
$
\rho=U_\rho.
$ 
\end{proposition}

From Proposition~\ref{prop1}, varying $\rho$ over its feasible range provides a way to examine how different configurations of latent compliance behavior affect the identification of both the ATE and LATE, with monotonicity included as a special case.

\emph{Third}, restrictions on $\rho$ can incorporate substantive knowledge about latent compliance behavior.  As discussed by \citet{wu2026quantifying} and \citet{Bodik-etal2025}, domain experts may be able to specify a plausible range for $\rho$, or at least its sign, based on subject-matter knowledge or auxiliary data. In particular, a positive correlation is plausible in many IV applications. For example, in studies of returns to schooling, let $Z$ denote college proximity and $T$  indicate college attendance. Individuals with a stronger underlying propensity to attend college may be more likely to enroll under either instrument assignment, leading to a positive association between $T(0)$ and $T(1)$. 
A similar argument applies in medical studies,
where $Z$ denotes a physician's treatment recommendation and $T$ denotes the treatment actually received. Patients with greater access to treatment or stronger adherence may be more likely to receive treatment under either assignment, again suggesting a positive association between the two potential treatment statuses. 

In contrast, monotonicity may be questionable in some empirical settings~\citep{Diegert2026}. For instance, in studies of
returns to schooling, college proximity may affect not only
the opportunity cost of attending college but also the cost
of enrollment, including the psychological effort required
and tuition expenses. Such competing mechanisms may generate all four principal strata~\citep{kedagni2024robust}.
In these settings, restricting the cross-world correlation provides a more flexible way to incorporate prior information about compliance behavior while allowing defiers to exist. 

To investigate how restrictions on the range of $\rho$ affect the sharp bounds for the ATE and LATE, we introduce the following Assumption  \ref{assump4}. 
\begin{assumption}[Cross-World Correlation Restriction] \label{assump4}
$\rho \in [\rho_l, \rho_u]$, where $\rho_l$ and $\rho_u$ are two prespecified thresholds satisfying $L_\rho \leq \rho_l \leq \rho_u \leq U_\rho$.  
\end{assumption}

The restriction $L_\rho \leq \rho_l \leq \rho_u \leq U_\rho$ is imposed without loss of generality,  since any prespecified range extending beyond the feasible range $[L_\rho,U_\rho]$ can be replaced
by its intersection with that range.   
When $\rho_l = L_{\rho}$ and $\rho_u = U_{\rho}$, Assumption~\ref{assump4} imposes no additional restrictions beyond Assumption~\ref{assump1}. In contrast, when $\rho_l > L_{\rho}$ or $\rho_u < U_{\rho}$, 
it restricts the proportions of the four principal strata and may therefore provide additional information for identifying the ATE and LATE.  
Table~\ref{tab:identification-roadmap} summarizes the identification results under
three different set of assumptions.

\begin{table}[H]
\centering
\caption{Main Results}
\label{tab:identification-roadmap}
\begingroup
\small
\renewcommand{\arraystretch}{1.15}
\setlength{\tabcolsep}{6pt}
\begin{tabular}{@{}l p{4.0cm} p{4.4cm} p{4.2cm}@{}}
\toprule
\textbf{}
 & \textbf{Basic Setup}
 & \textbf{With additional}
 & \textbf{With additional} \\ 
\textbf{Estimand}
 & \textbf{Assumptions \ref{assump1}--\ref{assump2}}
 & \textbf{Assumption \ref{assump3}}
 & \textbf{Assumption \ref{assump4}} 
 \\
\midrule
ATE
 & Balke--Pearl Bounds
 & Unchanged
 & Improved  \\
\addlinespace[0.5em]
LATE
 & Sharp Bounds  
   & Point Identified 
 & Improved 
 \\
\bottomrule
\end{tabular}
\endgroup
\par\vspace{0.5em}
\begin{minipage}{\textwidth}
\footnotesize
\textit{Notes:} 
``Unchanged'' indicates that the additional assumption does not tighten the identification bounds relative to the basic setup under Assumptions~\ref{assump1}--\ref{assump2}, whereas ``improved'' indicates that the additional assumption can yield tighter bounds. 
Moreover, the Wald ratio, which point-identifies the LATE under the additional Assumption~\ref{assump3}, always lies within the baseline LATE identified set under Assumptions~\ref{assump1}--\ref{assump2}, but may fall outside the LATE identified set obtained under the additional Assumption~\ref{assump4}.
\end{minipage}
\end{table}

In Subsections \ref{sec3-2} and \ref{sec3-3}, 
we develop the corresponding
sharp identification results for the ATE and LATE, respectively.

\subsection{Sharp Bounds for the ATE}  \label{sec3-2}

To investigate how the sharp identified set for the ATE changes with the correlation coefficient $\rho$, we first derive sharp bounds for the ATE at a fixed value of $\rho$. At a given $\rho \in [L_\rho, U_\rho]$, equation \eqref{eq:stratum-proportions-rho} identifies the proportions of the four principal strata. 
Nevertheless, the ATE remains partially identified because the observed data do not fully identify the potential outcome distributions within these strata. Theorem~\ref{thm:ate-fixed-rho} derives closed-form sharp bounds for the ATE as functions of \(\rho\), expressed in terms of the cell probabilities $\{p_{yt|z}: y,t,z\in\{0,1\}\}$ and the proportion of defiers, \(\pi_d(\rho)\).

\begin{theorem}[Sharp Bounds for the ATE at a Fixed Correlation Coefficient]
\label{thm:ate-fixed-rho}
Suppose that Assumptions~\ref{assump1}--\ref{assump2} hold, and Assumption~\ref{assump4} is imposed with $\rho = \rho_l = \rho_u$. Then, the sharp identified set for the ATE is
 $[L_\tau(\rho), U_\tau(\rho)],$  
where  
\begin{equation*}
L_\tau(\rho) 
= \max \begin{pmatrix}
  -p_{10|0}-p_{01|0}\\
  -p_{10|1}-p_{01|1}\\
  p_{00|0}+p_{11|1}-1\\
  p_{11|0}+p_{00|1}-1\\
  \pi_d(\rho) - p_{00|1} - p_{10|1} + p_{00|0}-p_{01|0}-p_{01|1}\\
  \pi_d(\rho) - p_{10|1} -p_{01|0} - p_{01|1}\\
    \pi_d(\rho)  - p_{10|1} -p_{10|0}-p_{01|0}\\
      \pi_d(\rho) - p_{00|1} + p_{00|0}- 2p_{10|1} - p_{01|1}\\  
  2 \pi_d(\rho)  - p_{00|1}  - 2  p_{10|1} + p_{00|0} -p_{01|0}-p_{01|1}
\end{pmatrix},
\end{equation*}
and
\begin{equation*}
U_\tau(\rho)
=\min 
\begin{pmatrix}
  p_{00|1}+p_{11|1}\\
  p_{00|0}+p_{11|0}\\
  1-p_{10|0}-p_{01|1}\\
  1-p_{01|0}-p_{10|1}\\
   - \pi_d(\rho) + p_{00|1} + p_{10|1}  + p_{11|0} + p_{11|1}-p_{10|0}\\
     - \pi_d(\rho) + p_{00|1} + p_{11|0}+ p_{11|1} \\
     - \pi_d(\rho) + p_{00|1}  + p_{00|0}+p_{11|0} \\
     - \pi_d(\rho) + 2p_{00|1} + p_{10|1}   +p_{11|1}-p_{10|0}  \\ 
  -2 \pi_d(\rho) + 2 p_{00|1} +  p_{10|1}  +p_{11|0}+p_{11|1}-p_{10|0}
\end{pmatrix}.
\end{equation*}
\end{theorem}

The first four terms in each bound in Theorem \ref{thm:ate-fixed-rho} are invariant to \(\rho\), whereas the remaining terms depend on \(\rho\) only through the proportion of defiers, \(\pi_d(\rho)\). Since \(\pi_d(\rho)\) decreases with \(\rho\), the lower bound \(L_\tau(\rho)\) decreases and the upper bound \(U_\tau(\rho)\) increases as \(\rho\) increases. Consequently, a smaller value of \(\rho\) yields a narrower identified set for the ATE. 
This pattern can be understood from how \(\rho\) affects the composition of the principal strata. As \(\rho\) decreases, the proportions of never-takers and always-takers decrease, whereas those of compliers and defiers increase. For never-takers, the potential outcome under treatment is never observed; conversely, for always-takers, the potential outcome under control is never observed. These strata therefore contribute greater uncertainty to the identification of the ATE. By contrast, treatment receipt changes with the instrument for compliers and defiers, allowing the observed outcomes under different instrument assignments to provide information about both potential outcomes. This explains why a smaller \(\rho\), which reduces the proportions of never-takers and always-takers, leads to tighter ATE bounds.

In practice, specifying a plausible range for \(\rho\) may be more defensible than specifying its exact value, as discussed in Subsection~\ref{sec3-1}. The monotonicity of the ATE bounds with respect to \(\rho\) allows us to obtain the sharp identified set under such an interval restriction directly. 
\begin{corollary}
\label{cor:ate-range}
Suppose that Assumptions \ref{assump1}, \ref{assump2}, and \ref{assump4}
hold with $\rho_l \neq \rho_u$. 
Then, the sharp identified set for the ATE is $[L_\tau(\rho_u), U_\tau(\rho_u)]$, where $L_\tau(\cdot)$ and $U_\tau(\cdot)$ are defined in  Theorem~\ref{thm:ate-fixed-rho}.  
\end{corollary}

At the upper endpoint $\rho=U_\rho$, the proportion of defiers is zero, corresponding to the monotonicity assumption. Substituting $\rho=U_\rho$ into Theorem~\ref{thm:ate-fixed-rho} recovers the Balke--Pearl bounds; see Supplementary Material S1 for details. This result is consistent with previous findings that imposing monotonicity does not tighten the Balke--Pearl bounds for the ATE relative to those obtained under Assumptions~\ref{assump1}--\ref{assump2} alone~\citep{Richardson2014,BaiEtAl2026}. 

Since both bounds in Theorem~\ref{thm:ate-fixed-rho} are monotone in $\rho$ and attain their widest values at $\rho=U_\rho$, the Balke--Pearl bounds at $\rho=U_\rho$ provide an outer bound for the identified sets over the entire feasible range of $\rho$. Consequently, restricting $\rho$ to an interval with upper endpoint $\rho_u<U_\rho$ can potentially yield tighter bounds than the Balke--Pearl bounds at $\rho=U_\rho$. However, such a restriction does not necessarily yield a strict improvement for every feasible range. We establishe an explicit threshold that characterizes when the proposed bounds are strictly tighter than the Balke--Pearl bounds at $\rho=U_\rho$; see Subsection~\ref{sec3-improve} for details.

\subsection{Sharp Bounds and Point Identification of the LATE}  \label{sec3-3}
We now turn to the LATE. 
Theorem ~\ref{thm:late-fixed-rho} presents sharp identified set for the LATE.

\begin{theorem}[Sharp Bounds for the LATE at a Fixed Correlation Coefficient] \label{thm:late-fixed-rho} 
Suppose that Assumptions~\ref{assump1}--\ref{assump2} hold, and Assumption~\ref{assump4} is imposed with $\rho = \rho_l = \rho_u$.  Then, the sharp identified set for the LATE is $[L_{\tau_c}(\rho),U_{\tau_c}(\rho)],$  
where 
\begin{equation*} 
L_{\tau_c}(\rho) 
       =  \frac{1}{ \Delta_T + \pi_d(\rho)}  \max \begin{pmatrix}
        - p_{01|1} \\
        - p_{10|0} \\
        p_{00|0} - p_{00|1} - p_{01|1} \\
        p_{00|0} + p_{01|0} + p_{11|1} - 1 \\
        -\Delta_T - \pi_d(\rho)  \\
       \Delta_T + \pi_d(\rho)   - p_{10|0} - p_{01|1} \\
         -\Delta_T - \pi_d(\rho) + p_{00|0} - p_{00|1} \\
         -\Delta_T - \pi_d(\rho) + p_{11|1} - p_{11|0} \\
          - \pi_d(\rho) + \Delta_Y
       \end{pmatrix}, 
\end{equation*} 
and 
\begin{equation*} 
U_{\tau_c}(\rho) 
       =  \frac{1}{ \Delta_T + \pi_d(\rho) }  \min \begin{pmatrix}
        p_{00|0} \\
        p_{11|1} \\
        p_{00|0} + p_{01|0} - p_{01|1} \\
        p_{11|1} + p_{10|1} - p_{10|0} \\
      \Delta_T + \pi_d(\rho)  \\
       \Delta_T + \pi_d(\rho)  + p_{01|0} - p_{01|1} \\
       \Delta_T + \pi_d(\rho) + p_{10|1} - p_{10|0} \\
      -\Delta_T - \pi_d(\rho)  + p_{00|0} + p_{11|1} \\
     \pi_d(\rho) + \Delta_Y
       \end{pmatrix}. 
\end{equation*} 
\end{theorem}

The denominator in both bounds of Theorem~\ref{thm:late-fixed-rho} equals the proportion of compliers, since $\pi_c(\rho)=\pi_d(\rho)+\Delta_T$. 
The first four terms in the numerator of each bound depend only on the cell probabilities $\{p_{yt|z}: y,t,z\in\{0,1\}\}$, whereas the remaining five terms also depend on \(\rho\) through \(\pi_d(\rho)\).

At the boundary value $\rho=U_\rho$, which corresponds to the monotonicity assumption and implies $\pi_d(\rho)=0$, the bounds in Theorem~\ref{thm:late-fixed-rho} collapse to the Wald ratio $w = \Delta_Y/\Delta_T$. 
This follows because the ninth terms in both the lower and upper bounds reduce to \(w\) at this boundary value. 
When \(\rho<U_\rho\), defiers remain possible, and for any feasible value of \(\rho\), the ninth terms give the following outer bounds on the sharp LATE interval:  
\begin{align*}
  \frac{\Delta_Y-\pi_d(\rho)}{\Delta_T+\pi_d(\rho)}
  \leq L_{\tau_c}(\rho)\leq U_{\tau_c}(\rho)
  \leq\frac{\Delta_Y+\pi_d(\rho)}{\Delta_T+\pi_d(\rho)}.
\end{align*}
This inequality provides a sufficient condition for identifying the sign of the LATE: if $\pi_d(\rho)<|\Delta_Y|$, the sharp identified set excludes zero, and the LATE has the same sign as $\Delta_Y$.  

For a prespecified correlation restriction $\rho\in[\rho_l,\rho_u]$, where $\rho_l<\rho_u$, Theorem \ref{thm:late-fixed-rho} implies that the sharp lower and upper bounds for the LATE are obtained by minimizing $L_{\tau_c}(\rho)$ and maximizing $U_{\tau_c}(\rho)$, respectively, over the restricted range:
\begin{align*}
    L_{\tau_c} ={} & \min_{\rho \in [\rho_l,\rho_u]} L_{\tau_c}(\rho), \quad
    U_{\tau_c} ={} \max_{\rho \in [\rho_l,\rho_u]} U_{\tau_c}(\rho).
\end{align*}
Thus, restricting $\rho$ to a narrower range can potentially tighten the LATE bounds.


From Theorem~\ref{thm:late-fixed-rho}, both the LATE bounds, $L_{\tau_c}(\rho)$ and $U_{\tau_c}(\rho)$, are not necessarily monotone in $\rho$. Nevertheless, the patterns of variation of these bounds as $\rho$ changes can be characterized more precisely; see Remark \ref{rmk:late-shape} for more details.

\begin{remark}[Variation of the LATE bounds with $\rho$]
\label{rmk:late-shape} 
The variation of the LATE bounds with $\rho$ can be characterized through the transformation $v=1/\pi_c(\rho)$. 
Each term in each bound in Theorem~\ref{thm:late-fixed-rho} can be written as 
\[
  \frac{a+b\pi_c(\rho)}{\pi_c(\rho)}=av+b,
\]
where \(a\) and \(b\) do not depend on \(\rho\). 
It follows that the lower and upper bounds are, respectively, the maximum and minimum of finitely many affine functions of \(v\).  
Because $v$ increases with $\rho$, the variation of the LATE bounds can be characterized by how the bounds vary with $\rho$.   
Specifically, we show  (see Supplementary Material S3 for proofs) that the lower bound may first decrease and then increase, or decrease or increase monotonically, but cannot transition from increasing to decreasing. Conversely, the upper bound may first increase and then decrease, or decrease or increase monotonically, but cannot transition from decreasing to increasing. Consequently, increasing $\rho$ does not necessarily tighten or widen the LATE identified interval uniformly, and the identified intervals corresponding to different values of $\rho$ need not be nested. 
\end{remark}

Having characterized how the bounds vary with \(\rho\), we next examine the conditions under which the sharp LATE interval collapses to a single point. Let 
\[
  \mu_{t,c}
  =\mathbb{E}\!\left[Y(t)\mid T(0)=0,T(1)=1\right],
  \qquad t\in\{0,1\},
\]
and \(\tau_c=\mu_{1,c}-\mu_{0,c}\). 
The following theorem presents the necessary and sufficient conditions for point identification of $\mu_{1,c}$ at any feasible value of $\rho$.





\begin{theorem}[Point identification of $\mu_{1,c}$ at fixed $\rho$]
\label{thm3}
Suppose that Assumptions~\ref{assump1}--\ref{assump2} hold, and Assumption~\ref{assump4} is imposed with $\rho = \rho_l = \rho_u$. Then, for any feasible $\rho$, $\mu_{1,c}$ is point identified if and only if at least one of the following
conditions holds:
\begin{enumerate}[label=(\roman*)]
  \item at least one of the four observed probabilities
  $p_{01\mid0}$, $p_{11\mid0}$, $p_{01\mid1}$, and $p_{11\mid1}$
  is zero.
  \item $\pi_a(\rho)=0$.
  \item $\pi_d(\rho)
    =\max\!\left\{
      0,\,
      p_{01\mid0}-p_{01\mid1},\,
      p_{11\mid0}-p_{11\mid1}
    \right\}.$
\end{enumerate}
\end{theorem}

A symmetric characterization of point identification for $\mu_{0,c}$ is provided in Supplementary Material~S2. 
The classical result that the LATE is point identified by the Wald ratio under monotonicity follows as a special case of Theorem \ref{thm3}.
Specifically, when \(\rho=U_\rho\) is feasible, Lemma \ref{lem:34-compatibility} in Section \ref{sec3-improve} implies that \(\pi_d(\rho)=0\), thus satisfying Condition (iii) in Theorem \ref{thm3} and yielding point identification. 
In addition, the point identification of both \(\mu_{0,c}\) and \(\mu_{1,c}\) yields point identification of the LATE, as shown in Corollary \ref{coro2}.
\begin{corollary}
\label{coro2}
Under the same assumptions of Theorem~\ref{thm3}, the LATE $\tau_c=\mu_{1,c}-\mu_{0,c}$ is point identified if and only if both $\mu_{1,c}$ and $\mu_{0,c}$ are point identified.
\end{corollary}

Notably, Condition (i) in Theorem \ref{thm3} does not depend on \(\rho\) and can be readily verified in empirical applications; see Example 4 in Section \ref{sec4}. In contrast, Conditions (ii) and (iii) depend on \(\rho\) and are included for completeness. 
Given the practical relevance of Condition (i), we provide an explicit identification formula for \(\mu_{1,c}\) when this condition holds. 

\begin{corollary}
\label{coro3}
Suppose that Assumptions~\ref{assump1}--\ref{assump2} hold, Assumption~\ref{assump4} is imposed with $\rho = \rho_l = \rho_u$, and conditions (i) in Theorem \ref{thm3} holds. Then, for any feasible $\rho$, $\mu_{1,c}$ is identified as 
\begin{equation}\label{eq3}
  \mu_{1,c}
  =
  \begin{cases}
    1,
    & p_{01\mid 1}=0,\\[6pt]
    0,
    & p_{11\mid 1}=0,\\[6pt]
    \displaystyle
    1-\frac{p_{01\mid 1}}{\pi_c(\rho)},
    & p_{01\mid0}=0,\\[10pt]
    \displaystyle
    \frac{p_{11\mid 1}}{\pi_c(\rho)},
    & p_{11\mid0}=0.
  \end{cases}
\end{equation}
\end{corollary}

The first two cases in equation~\eqref{eq3} yield values of zero or one and do not depend on \(\rho\). In the last two cases, \(\rho\) enters only through the complier proportion \(\pi_c(\rho)\), so different feasible values of \(\rho\) may yield different identified values of \(\mu_{t,c}\).

Since the LATE bounds are not necessarily monotone in \(\rho\), the Wald ratio may lie outside the sharp identified set for the LATE under certain restrictions on \(\rho\). We derive an explicit threshold that characterizes the range of \(\rho\) for which this occurs; see Subsection~\ref{sec3-improve} for details.

\subsection{Correlation Thresholds for ATE Bound Tightening and Wald Ratio Breakdown}
\label{sec3-improve}

In this subsection, we characterize the ranges of \(\rho\) over which the bounds in Theorem~\ref {thm:ate-fixed-rho} are strictly tighter than those at the boundary \(\rho=U_\rho\), as well as the range of $\rho$ over which the Wald ratio lies outside the bounds in Theorem~\ref{thm:late-fixed-rho}.

Both the ATE bounds at $\rho=U_\rho$ and the point-identified Wald ratio for the LATE rely on monotonicity (Assumption \ref{assump3}). We therefore begin by examining whether monotonicity is compatible with the observed data distribution. 
Under Assumptions~\ref{assump1}--\ref{assump2}, compatibility with monotonicity is characterized by the observable inequalities in Lemma~\ref{lem:34-compatibility}~\citep{Kitagawa2015}.

\begin{lemma}[Compatibility with Monotonicity]
\label{lem:34-compatibility}
The observed data distribution is compatible with monotonicity under Assumptions \ref{assump1}--\ref{assump2} if and only if
\begin{equation}
\begin{gathered}
  p_{00\mid0}\geq p_{00\mid1},\qquad
  p_{10\mid0}\geq p_{10\mid1},\\
  p_{01\mid1}\geq p_{01\mid0},\qquad
  p_{11\mid1}\geq p_{11\mid0}.
\end{gathered}
\label{eq:app34-monotonicity-compatibility}
\end{equation}
\end{lemma}

Lemma~\ref{lem:34-compatibility} provides a necessary and sufficient condition for Assumption~\ref{assump3} to be compatible with the observed data distribution under Assumptions~\ref{assump1}--\ref{assump2}. 
If any of these inequalities is violated, no joint distribution of the potential treatments and outcomes can satisfy monotonicity while reproducing the observed data distribution under Assumptions~\ref{assump1}--\ref{assump2}. In this case, comparing the proposed bounds with their monotonicity-based counterparts is not meaningful.
When these inequalities hold, the Balke--Pearl bounds for the ATE under Assumptions~\ref{assump1}--\ref{assump2} coincide with those under Assumptions \ref{assump1}--\ref{assump3}~\citep{BaiEtAl2026,kitagawa2021identification}. 

Next, we present the correlation threshold below which the sharp identified set in Theorem~\ref{thm:ate-fixed-rho} is a proper subset of those at the boundary $\rho = U_{\rho}$. 
\begin{proposition}[Correlation Thresholds for Tightening Balke–Pearl Bounds]
\label{prop:ate-refinement} 
Suppose that Assumptions~\ref{assump1}--\ref{assump2} and the observable restrictions \eqref{eq:app34-monotonicity-compatibility} hold. Then
    \[  [L_\tau(\rho), U_\tau(\rho)] \subset  [L_\tau(U_\rho), U_\tau(U_\rho)]    \quad\Longleftrightarrow\quad   \rho<\bar{\rho}_\tau,  \]
where 
\[
    \bar{\rho}_\tau = 
    U_\rho - \frac{d_{\tau}}{\gamma}, \quad d_{\tau} = \min\{p_{00\mid1},p_{01\mid0},p_{10\mid1},p_{11\mid0}\}.
\]
\end{proposition}

Proposition \ref{prop:ate-refinement}
implies that, when $\rho < \bar \rho_\tau$, restricting the cross-world correlation provides more informative bounds for the ATE than imposing monotonicity under 
 Assumptions \ref{assump1}--\ref{assump2}.  
Moreover, noting that \(\pi_d(\rho)=\lambda_0(1-\lambda_1)-\gamma\rho=\gamma(U_\rho-\rho)\),  the condition \(\rho<\bar \rho_\tau\) is equivalent to \(\pi_d(\rho)>d_\tau\), 
Thus, $d_\tau$ represents the minimum proportion of defiers required for the proposed bounds in Theorem~\ref{thm:ate-fixed-rho} to strictly improve upon those under monotonicity. 

We then characterize the correlation threshold below which  the Wald ratio falls outside the sharp identified set in Theorem~\ref{thm:late-fixed-rho}.  

\begin{proposition}[Compatibility of Wald Ratio]  \label{prop:wald-compatibility}  
Suppose that Assumptions~\ref{assump1}--\ref{assump2} and the observable restrictions \eqref{eq:app34-monotonicity-compatibility} hold. Then  
\[
   w \notin [L_{\tau_c}(\rho),U_{\tau_c}(\rho)]
  \quad\Longleftrightarrow\quad \rho <  \bar{\rho}_{\tau_c},
\]
where \begin{align*}
  \bar{\rho}_{\tau_c} ={}& U_\rho-\frac{d_{\tau_c}}{\gamma},\,\, d_{\tau_c} = \min\biggl\{ \lambda_0, 1 - \lambda_1, 
    \dfrac{p_{00\mid1}+p_{11\mid0}}{1+ w},
    \dfrac{p_{01\mid0}+p_{10\mid1}}{1- w}, \dfrac{\min\{p_{00\mid1},p_{11\mid0}\}}{(w)_+},
    \dfrac{\min\{p_{01\mid0},p_{10\mid1}\}}{(- w)_+}\biggr\},
\end{align*}
and $x_+=\max\{x,0\}$. 
Here, any ratio with zero denominator, including $0/0$, is defined as $+\infty$.
\end{proposition}

In Proposition \ref{prop:wald-compatibility}, \(\bar{\rho}_{\tau_c}\) represents the breakdown point below which the Wald ratio is no longer contained in the sharp identified set for the LATE, while \(d_{\tau_c}\) represents the largest proportion of defiers for which the Wald ratio remains within the identified set.
 We further show that $\bar \rho_\tau$ and $\bar \rho_{\tau_c}$  are comparable, as shown in Proposition \ref{prop:threshold-comparison}.  

\begin{proposition}[Comparison of the Thresholds]
\label{prop:threshold-comparison}
The thresholds $\bar \rho_\tau$ and $\bar \rho_{\tau_c}$ satisfy 
\[
  \bar \rho_{\tau_c} \leq \bar \rho_\tau, 
\]
and their difference satisfies 
\[ 
   \bar \rho_\tau - \bar \rho_{\tau_c}  \geq \frac{d_\tau}{\gamma} \frac{1-|w|}{ 1 + |w|}.
  \]  
In particular, if $d_\tau>0$ and $|w|<1$, then $\bar \rho_{\tau_c} <  \bar \rho_{\tau}$.  
\end{proposition}

Proposition \ref{prop:threshold-comparison} shows that 
$\bar \rho_{\tau_c} \leq \bar \rho_\tau$ always holds
 and provides a lower bound on the difference $\bar \rho_\tau - \bar \rho_{\tau_c}$.   
 In addition, we derive the necessary and sufficient conditions for the strict inequality $ \bar \rho_{\tau_c}< \bar \rho_\tau $ in Supplementary Material S2.  
Based on Propositions \ref{prop:ate-refinement}--\ref{prop:threshold-comparison}, three cases arise. 
First, when \(\rho<\bar{\rho}_{\tau_c}\), the Wald ratio is no longer a valid identification formula for the LATE, while the proposed ATE bounds in Theorem~\ref{thm:ate-fixed-rho} strictly improve upon the Balke--Pearl bounds under Assumptions~\ref{assump1}--\ref{assump3}. Second, when \(\bar{\rho}_{\tau_c}\leq\rho<\bar{\rho}_\tau\), the Wald ratio remains a valid identification formula for the LATE, and the proposed ATE bounds continue to strictly improve upon the Balke--Pearl bounds. Finally, when \(\rho\geq\bar{\rho}_\tau\), the Wald ratio remains a valid identification formula for the LATE, but the proposed ATE bounds no longer improve upon the Balke--Pearl bounds.

\section{Empirical Application} \label{sec4}

We illustrate the proposed method using four randomized studies with imperfect compliance. We demonstrate how additional restrictions on the cross-world correlation can tighten the bounds on the ATE and LATE, as well as assess
and assess whether the Wald ratio is compatible with the observed data under these restrictions.  
Specifically, based on the theoretical results in Section~\ref{sec3}, for each real-world dataset, we organize the empirical analysis into the following steps: 
\begin{itemize} 
    \item[] \emph{Step 1 (Preliminary Assessment and Threshold Calculation)}: 
    Calculate \(\hat\Delta_T\). If \(\hat\Delta_T<0\), swap the labels \(Z=0\) and \(Z=1\) to ensure that \(\hat\Delta_T>0\). 
    Then, calculate \(\hat L_\rho\), \(\hat U_\rho\), and the empirical 
    probabilities \(\hat p_{yt|z}\), and use these estimates to assess 
    whether monotonicity compatibility condition in 
    \eqref{eq:app34-monotonicity-compatibility} is  satisfied. 
    If so, further calculate the threshold estimates 
    \(\hat{\bar\rho}_\tau\) and \(\hat{\bar\rho}_{\tau_c}\).
    
    \item[] \emph{Step 2 (Baseline Estimation)}: 
    Estimate the baseline bounds, including the sharp bounds for the 
    ATE and LATE under Assumptions~\ref{assump1}--\ref{assump2}, and the Wald ratio for the LATE.  
    If the monotonicity compatibility condition in \eqref{eq:app34-monotonicity-compatibility} is satisfied, additionally estimate the sharp ATE bounds 
    under Assumptions~\ref{assump1}--\ref{assump3}.
    These estimates serve as benchmarks for comparison. 
    We denote the estimated Wald ratio by \(\hat w\).
    
     \item[] \emph{Step 3 (Correlation-Restricted Bounds and Comparisons)}: Over a prespecified grid of values of \(\rho\), estimate the sharp 
    bounds for the ATE and LATE given in Theorems~\ref{thm:ate-fixed-rho} 
    and~\ref{thm:late-fixed-rho}, respectively. 
    Denote the resulting estimates by 
    \(\hat L_\tau(\rho)\), \(\hat U_\tau(\rho)\), 
    \(\hat L_{\tau_c}(\rho)\), and \(\hat U_{\tau_c}(\rho)\), 
    respectively. Compare these correlation-restricted bounds with the baseline estimates obtained in Step 2.
\end{itemize}  
All analyses are conducted under Assumptions~\ref{assump1}--\ref{assump2}.

\paragraph{Example 1 (FLU).}
\citet{hirano2000} examined the effect of influenza vaccination on flu-related hospitalization using data from a randomized encouragement study in which physicians were randomly selected to receive letters encouraging vaccination of their patients. In the sample of 2,861 patients, $Z=1$ indicates that the patient's physician received the encouragement letter, and $Z=0$ indicates that no letter was received. For each patient, $T=1$ indicates influenza vaccination and $T=0$ no vaccination; $Y=1$ indicates a flu-related hospitalization during winter and $Y=0$ no such hospitalization. The encouragement letter is intended to affect hospitalization through vaccination, while its randomized allocation supports independence from potential vaccination decisions and hospitalization outcomes. Together, these design features lend plausibility to Assumptions~\ref{assump1}--\ref{assump2}. 

Patients whose physicians received the encouragement letter had higher vaccination uptake (30.77\% versus 18.93\%, \(\hat\Delta_T=0.118>0\)) and a lower flu-related hospitalization rate (7.81\% versus 9.29\%, \(\hat\Delta_Y=-0.015\)) than those whose physicians received no letter. The estimated lower and upper endpoints of the feasible range for \(\rho\) are \(\hat L_\rho=-0.322\) and \(\hat U_\rho=0.725\), respectively. By calculation, the empirical probabilities do not satisfy the monotonicity compatibility condition in \eqref{eq:app34-monotonicity-compatibility}, so we do not calculate \(\hat{\bar\rho}_\tau\) and \(\hat{\bar\rho}_{\tau_c}\). 
For the baseline estimates, the sharp ATE and LATE identified interval under Assumptions~\ref{assump1}--\ref{assump2} are $[-0.239,0.642]$ and $[-0.406,0.049]$, respectively. Both contain zero, so the lower hospitalization rate under encouragement does not identify the direction of either treatment effect under  Assumptions~\ref{assump1}--\ref{assump2}. 
The estimated Wald ratio is \(\hat w=-0.125\), indicating a protective effect among compliers under the additional monotonicity (Assumption~\ref{assump3}). 
For the proposed method, we calculate \(\hat L_\tau(\rho)\), \(\hat U_\tau(\rho)\), \(\hat L_{\tau_c}(\rho)\), and \(\hat U_{\tau_c}(\rho)\) 
over a grid of \(\rho\) values.  

\begin{figure}[htbp]
\centering
\includegraphics[width=\linewidth]{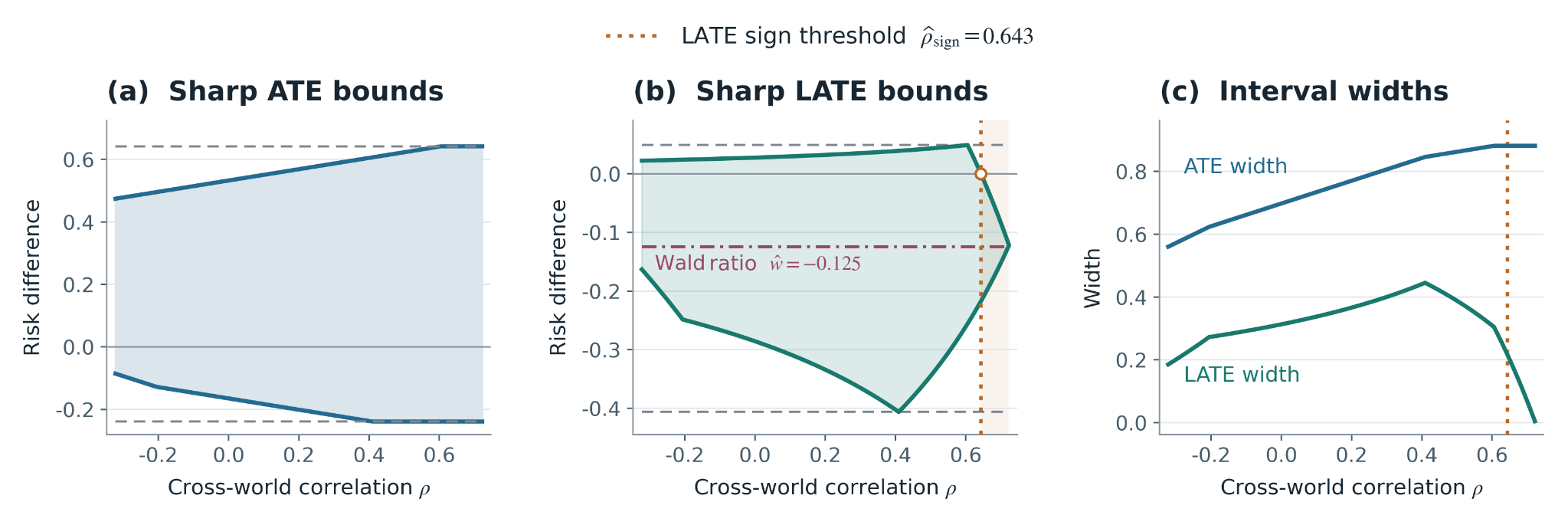}
\caption{FLU. (a) ATE bounds, where the blue solid lines represent the proposed sharp bounds over a range of $\rho$, and the gray dashed lines represent the Balke--Pearl bounds under Assumptions~\ref{assump1}--\ref{assump2};
(b) LATE bounds, where the green solid lines represent the proposed sharp bounds over a range of $\rho$, and the gray dashed lines represent the corresponding bounds under Assumptions~\ref{assump1}--\ref{assump2}. The purple horizontal line represents the Wald ratio, and the orange vertical line marks the threshold at which the  proposed upper bound for the LATE crosses zero; 
(c) Interval widths, where the blue and green solid lines represent the widths of the proposed sharp ATE and LATE bounds at each value of $\rho$, respectively. The orange vertical line marks the same LATE sign threshold as in (b).}
\label{fig:flu}
\end{figure}

From Figure~\ref{fig:flu}(a), as \(\rho\) increases, the ATE lower bound decreases to the Balke--Pearl lower bound of \(-0.239\) at \(\rho=0.409\) and remains constant thereafter; the upper bound increases to its Balke--Pearl upper bound of \(0.642\) at \(\rho=0.605\) and then remains constant. From Figure~\ref{fig:flu}(b), as \(\rho\) increases, the LATE lower bound first decreases to its minimum of \(-0.406\) at \(\rho=0.409\) and then increases, while the upper bound first increases to its maximum of \(0.049\) at \(\rho=0.605\) and then decreases. 
The upper bound crosses zero at \(\widehat\rho_{\mathrm{sign}}=0.643\), so the LATE is strictly negative for all \(\rho>\widehat\rho_{\mathrm{sign}}\). This threshold characterizes the degree of compliance information required to establish a protective complier effect while allowing for defiers.  
 Figure~\ref{fig:flu}(c) summarizes the corresponding interval widths: the ATE width increases from its minimum of \(0.560\) at \(\rho=\hat L_\rho=-0.322\) to the Balke--Pearl width of \(0.881\) at \(\rho=0.605\), remaining constant thereafter. The LATE width increases to its maximum of \(0.445\) at \(\rho=0.409\) and then decreases. 
\paragraph{Example 2 (IMPROVE).}
The IMPROVE trial evaluated whether an endovascular strategy improves 30-day survival compared with open repair for ruptured abdominal aortic aneurysm \citep{improve2014}. We use the cross-tabulation of 501 patients with confirmed rupture who underwent repair: $Z=1$ denotes assignment to the endovascular strategy and $Z=0$ assignment to open repair; $T=1$ denotes actual endovascular repair and $T=0$ open repair; $Y=1$ denotes survival at 30 days and $Y=0$ death within 30 days. Random allocation supports independence between assignment and potential repair decisions (Assumption~\ref{assump1}) and survival outcomes in the full trial population, and assignment in this study is assumed to affect 30-day survival only through the repair received  (Assumption~\ref{assump2}).
 
Patients assigned to the endovascular strategy had higher rates of endovascular repair (57.53\% versus 13.22\%, \(\hat\Delta_T=0.443>0\)) and 30-day survival (67.57\% versus 64.05\%).
The estimated feasible range for $\rho$ is \( [\hat L_\rho, \hat U_\rho]=[-0.454, 0.335]\). 
By calculation, the empirical probabilities satisfy the monotonicity compatibility condition in \eqref{eq:app34-monotonicity-compatibility}, and the estimated threshold \(\hat{\bar\rho}_\tau=0.138\) and the breakdown point \(\hat{\bar\rho}_{\tau_c}=\hat L_\rho = -0.454\). 
For the baseline estimates, the sharp ATE and LATE intervals under Assumptions~\ref{assump1}--\ref{assump2} are \([-0.260,0.297]\) and \([0.004,0.248]\), respectively. The ATE interval contains zero, whereas the LATE interval lies entirely above zero, indicating a positive survival effect among compliers. Under the additional Assumption~\ref{assump3}, the sharp ATE interval aligns with those under Assumptions \ref{assump1}--\ref{assump2}, and the Wald ratio identifies the LATE, with an estimated value of \(\hat w=0.079\). For the proposed method, we calculate \(\hat L_\tau(\rho)\), \(\hat U_\tau(\rho)\), \(\hat L_{\tau_c}(\rho)\), and \(\hat U_{\tau_c}(\rho)\) over a grid of \(\rho\) values. 

\begin{figure}[H]
\centering
\includegraphics[width=\linewidth]{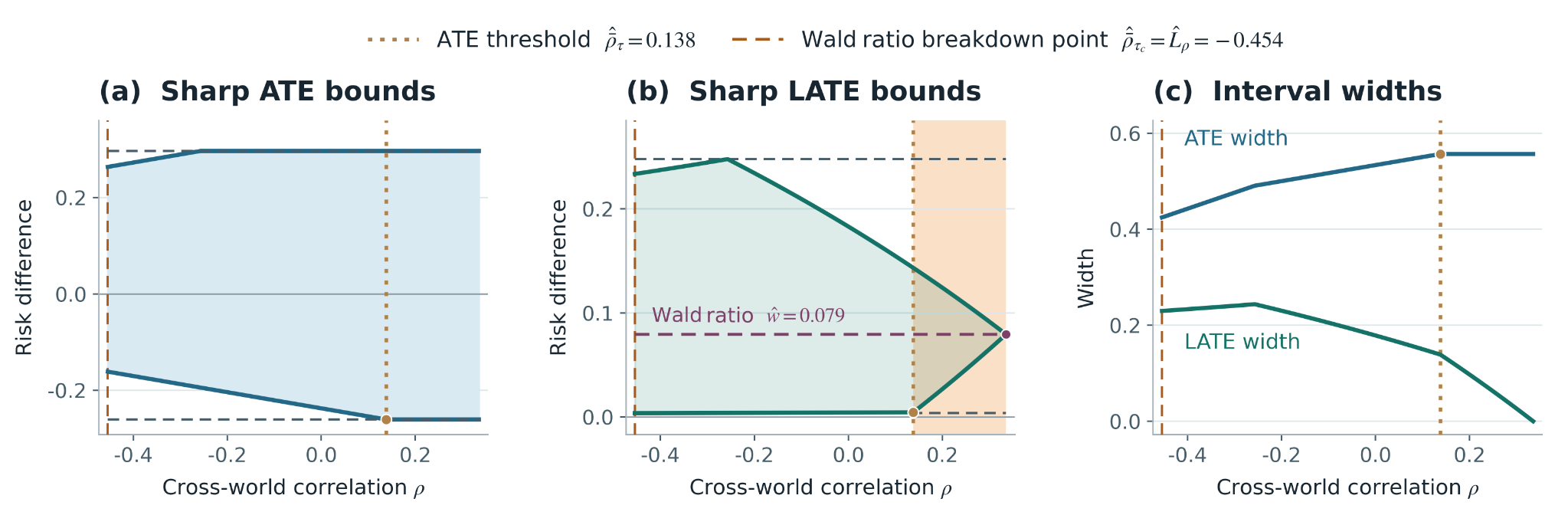}
\caption{IMPROVE. 
(a) ATE bounds, where the blue solid lines represent the proposed sharp bounds over a range of $\rho$, and the gray dashed lines represent the Balke--Pearl bounds under Assumptions~\ref{assump1}--\ref{assump2};
(b) LATE bounds, where the green solid lines represent the  proposed sharp bounds over a range of $\rho$, the gray dashed lines represent the corresponding bounds under Assumptions~\ref{assump1}--\ref{assump2}, and the purple horizontal line represents the Wald ratio; 
(c) Interval widths, where the blue and green solid lines represent the widths of the proposed sharp ATE and LATE bounds at each value of $\rho$, respectively.
In all three panels, the orange vertical lines mark the threshold $\hat{\bar \rho}_\tau = 0.138$ and the breakdown point  $\hat{\bar \rho}_{\tau_c} = -0.454$.
}
\label{fig:improve}
\end{figure}

From Figure~\ref{fig:improve}(a), as \(\rho\) increases, the ATE upper bound rises to the Balke--Pearl upper bound of \(0.297\) at \(\rho=-0.257\) and remains constant thereafter. The ATE lower bound decreases to its Balke--Pearl lower bound of \(-0.260\) at the threshold \(\hat{\bar\rho}_\tau = 0.138\) and then remains constant.
From  Figure~\ref{fig:improve}(b), 
the LATE upper bound first rises to \(0.248\) at \(\rho=-0.257\) and then decreases, while the lower bound increases slowly until \(\rho=0.138\) and more rapidly thereafter. Both bounds converge to the Wald estimate, \(\hat w=0.079\), at \(\rho=\hat U_\rho=0.335\). Since the breakdown point satisfies \(\hat{\bar\rho}_{\tau_c}=\hat L_\rho=-0.454\), the Wald ratio remains within the sharp identified set for the LATE throughout the feasible range.
 Figure~\ref{fig:improve}(c) shows that the ATE width increases from its minimum of \(0.425\) at \(\rho=\hat L_\rho=-0.454\) to \(0.557\) at \(\rho=0.138\), after which it remains constant. The LATE width first increases to its maximum of \(0.244\) at \(\rho=-0.257\) and then decreases to zero at \(\rho=\hat U_\rho=0.335\). 
Over the correlation range \(\rho\in[\hat{\bar\rho}_\tau,\hat U_\rho]=[0.138,0.335]\), the ATE interval remains equal to the Balke--Pearl interval. The estimated sharp LATE interval over the same range is \([0.005,0.143]\), with a width of \(0.139\), representing a \(43.11\%\) reduction relative to the baseline width of \(0.244\) under Assumptions~\ref{assump1}--\ref{assump2}.

\paragraph{Example 3 (Beam).}
The randomized encouragement experiment of \citet{beam2016job} examined whether attending a job fair affected subsequent employment in the rural Philippines. 
In the analytic sample of 220 participants, $Z=1$ indicates that the participant was offered a restaurant voucher conditional on attending the job fair, whereas $Z=0$ indicates that no such offer was made; $T=1$ denotes job-fair attendance and $T=0$ denotes nonattendance; and $Y=1$ denotes self-reported informal-sector employment approximately ten months after the fair, whereas $Y=0$ denotes any other observed employment status. The study randomized the restaurant voucher offer, ensuring that assignment was independent of potential attendance decisions and employment outcomes. Moreover, the voucher offer served as an encouragement to attend the job fair, and its effect on subsequent employment was expected to operate primarily through job-fair attendance. These design features provide support for Assumptions
Assumptions~\ref{assump1}--\ref{assump2}. 

The voucher group had higher rates of attendance (48.05\% versus 16.08\%, \(\hat\Delta_T=0.320>0\)) and informal-sector employment (18.18\% versus 11.19\%).
The estimated  feasible range for $\rho$ is \([\hat L_\rho,\hat U_\rho] = [-0.421,0.455]\). By calculation, the empirical probabilities satisfy the monotonicity compatibility condition in \eqref{eq:app34-monotonicity-compatibility}, and the estimated threshold \(\hat{\bar\rho}_\tau=0.417\) and the breakdown point \(\hat{\bar\rho}_{\tau_c} = 0.281\). 
For the baseline estimates, the sharp ATE and LATE intervals under Assumptions~\ref{assump1}--\ref{assump2}  are \([-0.188,0.493]\) and \([-0.080,0.236]\), respectively. Both intervals contain zero. 
Under the additional Assumption~\ref{assump3}, the sharp ATE interval coincides with that under Assumptions \ref{assump1}--\ref{assump2}, and  the Wald ratio identifies the LATE, with an estimated value of \(\hat w=0.2188\). 
 For the proposed method, we calculate \(\hat L_\tau(\rho)\), \(\hat U_\tau(\rho)\), \(\hat L_{\tau_c}(\rho)\), and \(\hat U_{\tau_c}(\rho)\) over a grid of \(\rho\) values. 

\begin{figure}[H]
\centering
\includegraphics[width=\linewidth]{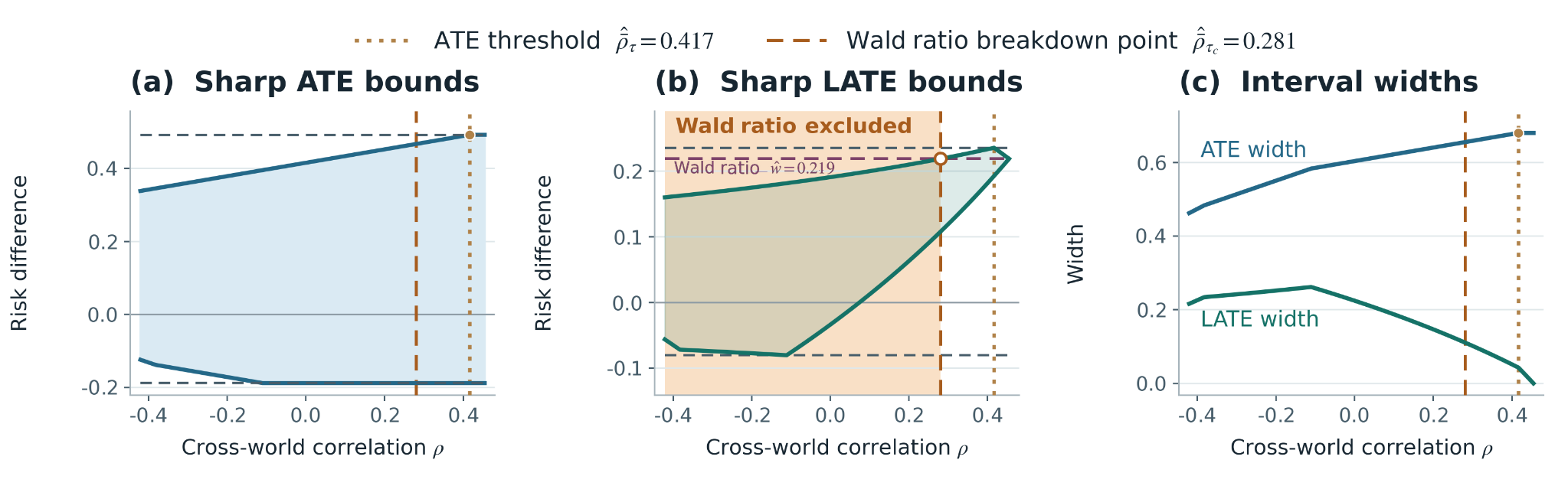}
\caption{Beam. 
(a) ATE bounds, where the blue solid lines represent the proposed sharp bounds over a range of $\rho$, and the gray dashed lines represent the Balke--Pearl bounds under Assumptions~\ref{assump1}--\ref{assump2};  
(b) LATE bounds, where the blue solid lines represent the  proposed sharp bounds over a range of $\rho$, the gray dashed lines represent the corresponding bounds under Assumptions~\ref{assump1}--\ref{assump2}, and the purple horizontal line represents the Wald ratio; 
(c) Interval widths, where the blue and green solid lines represent the widths of the proposed sharp ATE and LATE bounds at each value of $\rho$, respectively. In all three panels, the orange vertical lines mark the threshold $\hat{\bar \rho}_\tau = 0.417$ and the breakdown point  $\hat{\bar \rho}_{\tau_c} = 0.281$.}
\label{fig:wald}
\end{figure}

From Figure~\ref{fig:wald}(a), as \(\rho\) increases, the ATE lower bound decreases until it reaches the Balke--Pearl lower bound  \(\rho=-0.111\), while the ATE upper bound increases until it reaches the Balke--Pearl upper bound at the threshold \(\hat{\bar\rho}_\tau = 0.417\).  
Figure~\ref{fig:wald}(b) shows that the LATE lower bound reaches its minimum at \(\rho=-0.111\) and then rises.
The upper bound increases to its maximum at \(\hat{\bar\rho}_\tau=0.417\) and then decreases, until both bounds meet at the Wald ratio \(\hat w\) when \(\rho=\hat U_\rho = 0.455\). The orange shading indicates the feasible range of correlations below \(\hat{\bar\rho}_{\tau_c} = 0.281\), over which \(\hat w\) is excluded from the sharp identified set for the LATE.
Figure~\ref{fig:wald}(c) shows that 
the ATE interval widens to a maximum width of \(0.680\) at \(\hat{\bar\rho}_\tau=0.417\) and then levels off. The LATE interval widens to a maximum width of \(0.262\) at \(\rho=-0.111\) and then narrows to a point at \(\hat U_\rho\).

The results can be summarized by dividing the range of \(\rho\) into the following intervals: 
(a) When \(\rho\in[-0.421,0.281)\): The estimated sharp ATE and LATE intervals are  strictly narrower than their respective sharp intervals under Assumptions~\ref{assump1}--\ref{assump2}. Moreover, the Wald estimate exceeds the LATE upper bound and therefore cannot represent a valid estimate for the LATE;  
   (b) When \(\rho\in[0.281, 0.417)\):  Restricting \(\rho\) tightens the ATE bounds and allows the Wald ratio to fall within the sharp LATE interval; (c)
 When \(\rho\in[0.417, 0.455]\): Further restricting \(\rho\) does not tighten the ATE bounds, and the sharp LATE interval continues to contain the Wald ratio.

\paragraph{Example 4 (Head).}
The trial of \citet{head2002randomized} compared epidural and patient-controlled intravenous opioid analgesia during labor in women with severe preeclampsia. We use the 116-participant cross-tabulation reported by \citet{zhou2023}. For each participant , \(Z=1\) and \(Z=0\) denote assignment to epidural and intravenous analgesia, respectively. 
\(T=1\) and \(T=0\) denote receipt of epidural and intravenous analgesia, respectively, while \(Y=1\) and \(Y=0\) denote cesarean delivery and no cesarean delivery, respectively.
Random assignment to epidural or intravenous analgesia ensures independence between assignment and potential treatment receipt and cesarean delivery outcomes. Moreover, because assignment specifies the planned regimen for labor pain relief, any effect on cesarean delivery is expected to operate only through actual epidural receipt. Together, these considerations support Assumptions~\ref{assump1}--\ref{assump2}.

In the epidural and intravenous assignment arms, respectively, the rates of epidural receipt were \(53/56\) and \(2/60\) ($\hat\Delta_T=0.913>0$), and the cesarean delivery rates were \(10/56\) and \(7/60\). 
The estimated feasible range for $\rho$ is \([\hat L_\rho,\hat U_\rho] = [-0.781,0.044]\). By calculation, the empirical probabilities satisfy the monotonicity compatibility condition in \eqref{eq:app34-monotonicity-compatibility}, and the estimated thresholds  \(\hat{\bar\rho}_\tau=\hat{\bar\rho}_{\tau_c}=\hat U_\rho\). 
For the baseline estimates, the sharp ATE and LATE intervals under Assumptions~\ref{assump1}--\ref{assump2}  are \([0.029,0.112]\) and \([0.065,0.069]\), respectively. Both intervals indicate positive average treatment effects on cesarean delivery in the trial population and  compliers.
Under the additional Assumption~\ref{assump3}, the sharp ATE interval coincides with that under Assumptions \ref{assump1}--\ref{assump2}, and  the Wald ratio identifies the LATE, with an estimated value of \(\hat w=0.068\). 
 For the proposed method, we calculate \(\hat L_\tau(\rho)\), \(\hat U_\tau(\rho)\), \(\hat L_{\tau_c}(\rho)\), and \(\hat U_{\tau_c}(\rho)\) over a grid of \(\rho\) values.

\begin{figure}[H]
\centering
\IfFileExists{figure/fig4_head.png}{%
  \includegraphics[width=\linewidth]{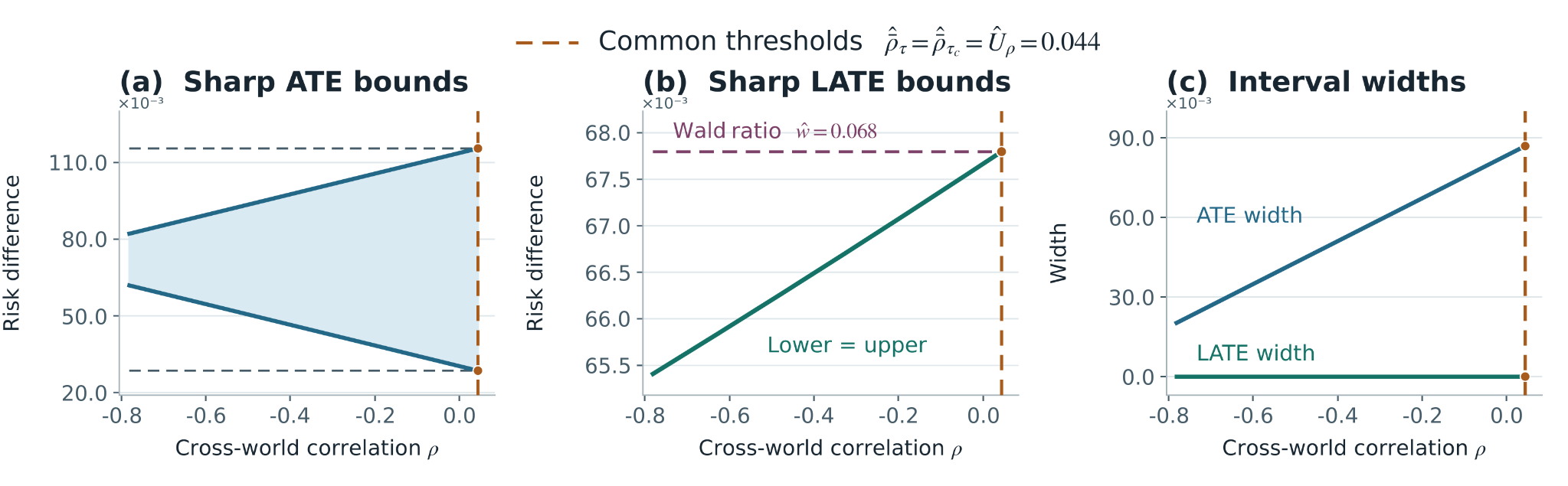}%
}{%
  \includegraphics[width=\linewidth]{figure/fig4_head.png}%
}
\caption{Head. Panels (a) and (b) show the sharp ATE and LATE bounds over a range of \(\rho\). Panel (c) shows both interval widths. Gray dashed lines give the Balke--Pearl bounds, and the purple dashed line gives the Wald estimate. Orange vertical lines mark the common thresholds.}
\label{fig:head}
\end{figure}

Figure~\ref{fig:head}(a) shows that the sharp ATE interval lies strictly inside the Balke--Pearl interval at every feasible \(\rho<\hat U_\rho\) and equal that interval at \(\hat U_\rho\). 
Its width increases from \(0.020\) at \(\hat L_\rho\) to \(0.087\) at \(\hat U_\rho\) (Figure~\ref{fig:head}(c)). 
Figure~\ref{fig:head}(b) shows that the lower and upper bounds for the LATE coincide. This is because the empirical probabilities satisfy the point-identification condition \(\hat p_{11\mid0}=\hat p_{10\mid1}=0\) in Corollary \ref{coro2} and leads to  at each fixed feasible $\rho$, yielding
\[
  \hat L_{\tau_c}(\rho)=\hat U_{\tau_c}(\rho)
  =\frac{\hat\Delta_Y}{\hat\Delta_T+\hat{\pi}_d(\rho)},
\]
where \(\hat{\pi}_d(\rho)\) denotes the empirical defier proportion at a given \(\rho\). Consequently, fixing \(\rho\) point-identifies the LATE, so the sharp LATE interval has zero width (Figure~\ref{fig:head}(c)). As \(\rho\) increases, the LATE approaches the Wald ratio  \(\hat w\) from below and equals it at \(\rho=\hat U_\rho\). Thus, \(\hat w\) lies outside the sharp LATE identified set for every  \(\rho<\hat U_\rho\).


\section{Conclusion}  \label{sec5}

In this article, we develop a partial identification approach for the ATE and LATE in randomized experiments with imperfect compliance. Maintaining random assignment and the exclusion restriction, we characterize the dependence between an individual's potential treatment statuses under alternative instrument assignments through their cross-world correlation.
We derive explicit sharp bounds for the ATE and LATE at a fixed value of the cross-world correlation and extend the results to prespecified correlation ranges. This provides a unified approach to incorporating information about latent compliance behavior into partial identification. 
The theoretical results reveal that restrictions on the cross-world correlation have different implications for the partial identification of the ATE and LATE, while also providing a framework for assessing when the Wald ratio breaks down.  
Further research would be desirable to extend the proposed approach to multivalued assignments and treatments. Such extensions would broaden its applicability but require interpretable restrictions on more complex compliance structures. Existing identification and computational methods for discrete causal models may provide a foundation for these extensions~\citep{duarte2024automated,song2024categorical}. 

\bibliographystyle{plainnat}
\bibliography{ref}

\end{document}